\documentstyle[prl,aps,epsfig]{revtex}
\begin{document}

\def\mev{\hbox{\ MeV}}
 


 \twocolumn[\hsize\textwidth\columnwidth\hsize  
 \csname @twocolumnfalse\endcsname              

\title{Nuclear Multifragmentation in the Non-extensive Statistics - Canonical
Formulation}

\author{K.K. Gudima\dag\ddag, A.S. Parvan\ddag\S,  M. P{\l}oszajczak\dag 
~and V.D. Toneev\dag\S}
\address{\dag\ Grand Acc\'{e}l\'{e}rateur National d'Ions Lourds,
CEA/DSM -- CNRS/IN2P3, BP 5027, F-14076 Caen Cedex 05, France}
\address{\ddag\ Institute of Applied Physics, Moldova Academy of 
Sciences, MD-2028 Kishineu, Moldova}
\address{\S\  Bogoliubov Laboratory of Theoretical Physics,  
Joint Institute for Nuclear Research, 141980 Dubna, Russia}


\maketitle

\begin{abstract}
We apply the canonical quantum statistical model of nuclear 
multifragmentation generalized in the framework of recently proposed Tsallis
non-extensive thermostatistics for the description of nuclear multifragmentation
process. The test calculation in the system with $A=197$
nucleons show strong modification of the 'critical' behaviour associated with
the nuclear liquid-gas phase transition for small deviations from the 
conventional Boltzmann-Gibbs statistical mechanics.
\end{abstract}


 ]  

\narrowtext
Most of the fragmenting systems are characterized by
strongly off-equilibrium processes which cease due to dissipation. The
theoretical description of the fragmentation process depends on whether
the equilibrium has been reached before system starts fragmenting. If the
equilibrium is attained, then the thermodynamic models using different
statistical ensembles in a given fixed volume (freeze-out volume) can be 
applied. The ingredients specific for the considered phenomenology enter
through the definition of fragments sizes and (binding) energies, fragment 
internal excitation properties, system size, conserved quantities in this
process etc. The observable characteristics of the fragmenting system are employed
to fix certain features of, in general unknown, intermediate equilibrium state.
In nuclear physics, several models of this kind have been tried with
unquestionable success in describing the transitional phenomenon in
heavy ion collisions from the regime of particle evaporation at lower
excitation energies to the explosion at about 5 - 10 MeV/nucleon of the hot 
source accompanied by the copious production of the intermediate mass fragments
\cite{mult2,mult3,mult4}. The situation when the fragment production has
to be considered as an off-equilibrium process is described by various  
kinetic equations, mainly on the level of one-body distribution functions
\cite{kin1}. Here the statistical equilibrium is not assumed 
but the kinetic models in turn are plagued by the unsurmountable conceptual 
difficulties in the calculation of asymptotic, observable features of the 
fragments. As an attempt to overcome at least some of these difficulties in
both groups of models, in this work we extend the thermodynamic (canonical) 
model of the fragmentation in the framework of the recently proposed
thermostatistics \cite{tsallis} to include certain off-equilibrium correlations in the system.
The Tsallis' generalized statistical mechanics (TGSM), which provides the basis
for generating this new model, is based on an alternative definition for the
equilibrium entropy of a system whose $i$th microscopic state has probability
${\hat p}_i$ :
\begin{equation}
\label{eq1}
S_q=k\frac{1-\sum_{i}{\hat p}_{i}^{q}}{q-1}~~, \hspace{1cm} \sum_{i}{\hat p}_i=1
\end{equation}
and $q$ (entropic index) defines a particular statistics. Entropy $S_q$ 
has the usual properties of positivity, equiprobability, concavity and 
irreversibility, and preserves the Legendre transformations structure of 
thermodynamics. In the limit : $q\rightarrow 1$, one
obtains the usual Boltzmann-Gibbs formulation of the statistical mechanics. The
main difference between the Boltzmann-Gibbs formulation and the TGSM lies in
the nonadditivity of the entropy.
Indeed, for two independent subsystems $A$, $B$,
such that the probability of $A+B$ is factorized into : $p_{A+B}=p_Ap_B$, the
global entropy verifies :
\begin{equation}
\label{eq2}
S_q(A+B)=S_q(A)+S_q(B)+(1-q)S_q(A)S_q(B)
\end{equation}
TGSM provides a natural framework for the thermodynamical formalism of the 
anomalous diffusion and ubiquity of Levy distributions \cite{zanette}. 
Long-range correlations in the system, 
as appearing in the situation of the thermalization of a hot gas penetrating 
in a cold gas in the presence of long range interactions, is 
typical for $q>1$  \cite{waldeer}. In some cases, the entropic index $q$
in TGSM can
can be also related to the fluctuations of the temperature in the system 
\cite{wilk99}. Variety of the off-equilibrium situations which can be
accounted for within the TGSM make it useful as a basis for the generalization
of the thermodynamical fragmentation models and , in particular in addressing
the problem of the influence of these non-extensivity correlations on the
signatures of 'criticality' in finite systems. In the context of nuclear
multifragmentation, this is usually referred to as the signatures of liquid-gas
phase transition in small systems.
     
Our starting point is the canonical multifragmentation model \cite{pgt99}
with the recurrence equation method \cite{pgt99,mek93} 
which makes the model solvable without the Monte Carlo technique and
transparent to the physical assumptions and generalizations. Canonical ensemble
method in TGSM was discussed in \cite{tsallis98}. The main ingredient of 
the TGSM generalization of the canonical fragmentation model \cite{pgt99}
is the expression for fragment partition function :
\begin{eqnarray}
\label{eq3}
{\cal Z}_q{(s,t)} &=& \sum_{\vec{p}}[1+q_1\beta 
\varepsilon_{\vec{p}}{(s,t)}]^{-1/q_1} 
\end{eqnarray}
where $q_1\equiv q-1$, $s$ and $t$ are the fragment 
mass and charge respectively, 
and the fragment partition probability equals :
\begin{eqnarray}
\label{eq4}
{\hat p}_{\vec {p}}{(s,t)}=({\cal Z}_q{(s,t)})^{-1}
[1+q_1\beta\varepsilon_{\vec {p}}{(s,t)}]^{-1/q_1}
\end{eqnarray}
where
$\varepsilon_{\vec {p}}{(s,t)}=p^2/2m+U{(s,t)}$, $\beta \equiv 1/T$.
The internal energy $U$ includes the fragment binding energy, the 
excitation energy and the Coulomb interaction between fragments in the 
Wigner-Seitz approximation \cite{mult3}. 
Changing summation into the integration in (\ref{eq3}) one gets :
\begin{eqnarray}
\label{eq5}
{\cal Z}_q{(s,t)}&=&\frac{gV_f}{\lambda_T^{3}}
\Gamma\left( {q_1}^{-1}-3/2\right){\Gamma}^{-1}\left({q_1}^{-1}
\right)\times \nonumber \\ &\times &q_1^{-3/2} 
[1+q_1\beta U{(s,t)}]^{-\frac{1}{q_1}+3/2} ~ \ ,
\end{eqnarray}
where $g$ is the spin degeneracy factor, $V_f$ is the free volume 
and $\lambda_T =[(2\pi )/(mT)]^{1/2}$,
where $m$ is the mass of the fragment $(s,t)$. 
In the limit $q\rightarrow 1$, one recovers the familiar expression :
${\cal Z}_1 = gV_f{\lambda_T}^{-3} \exp{(-\beta U)}$.

Given the partition function,
the mean value of any quantity in TGSM is \cite{tsallis} :
\begin{eqnarray}
\label{eq6}
<{\cal O}>_q=\sum_{\vec p}{\cal O}_{\vec p}{\hat p}_{\vec p}^q
\end{eqnarray}
For the average energy of fragment $(s,t)$ one obtains :
\begin{eqnarray}
\label{eq7}
<{\varepsilon(s,t)}>_q=-\frac{\partial}{\partial \beta}\left(
\frac{1-[{\cal Z}_q(s,t)]^{-q_1}}{q_1}\right) ~ \ .
\end{eqnarray}

In the dilute gas approximation \cite{bdg97}, a partition function 
of the whole system   can be written as follows :
\begin{eqnarray}
\label{eq8}
{\cal Q}_{q}{(A,Z)} = \sum_{{\hat n} \in {\Pi}_{A,Z}}{\prod}_{s,t}
\frac{\left[ {\cal Z}_{q}{(s,t)}\right]^{N_{{\hat n}}{(s,t)}}}
{N_{{\hat n}}{(s,t)}!} ~ \ ,
\end{eqnarray}
where the sum runs over the ensemble $\Pi_{A,Z}$ of different partitions of 
$A$ and $Z$ of the  decaying system : $\{{\hat n}\}=\{N_{{\hat n}}{(1,0)}, 
N_{{\hat n}}{(1,1)},\dots,N_{{\hat n}}{(A,Z)}\}$ and 
$N_{{\hat n}}{(s,t)}$ is the number of fragments of mass $s$ and charge $t$ in
the partition ${{\hat n}}$. In this approximation, the recurrence 
relation technique \cite{pgt99} can be applied providing exact
expression for ${\cal Q}_{q}{(A,Z)}$ :
\begin{eqnarray}
\label{eq9}
&&{\cal Q}_{q}{(A,Z)} = \nonumber \\&=&\frac{1}{A}\sum_{s,t<s;s-t<A-Z}s{\cal Z}_{q}{(s,t)}{\cal
Q}_{q}{(A-s,Z-t)} 
\end{eqnarray}
These relations can now be conveniently used to calculate ensemble averaged 
characteristics. However, in order to ensure 
the proper normalization, it is better to work with generalized averages 
\cite{tsallis1} :
\begin{eqnarray}
\label{eq10}
{\ll{\cal O}\gg}_q=<{\cal O}>_q/<1>_q ~ \ . 
\end{eqnarray} 
These normalized mean values exhibit all convenient properties of the
original mean values (\ref{eq6}). Moreover, when the normalized mean values
(\ref{eq10}) are used, the TGSM can be reformulated in terms of ordinary 
linear mean values 
calculated for the renormalized entropic index : $q^*=1+(q-1)/q$. 
In particular, the total average energy of the system becomes :
\begin{eqnarray}
\label{eq11}
{\cal E}_{q}= 
\sum_{s,t}<N{(s,t)}>_{q^* A Z} <\varepsilon(s,t)>_{q^*}
\end{eqnarray}
where $<\varepsilon(s,t)>_{q^*}$ is given in (\ref{eq7}) and :
\begin{eqnarray}
\label{eq12}
<N(s,t)>_{q A Z}={\cal Z}_{q}{(s,t)}
\frac{{\cal Q}_{q}{(A-s,Z-t)}}{{\cal Q}_{q}{(A,Z)}} ~ . \ 
\end{eqnarray}
Analogously, the heat capacity ($=\partial {\cal E}_q/\partial T\mid_V$) is :
\begin{eqnarray}
\label{eq13}
C_V&=&{\beta}^2\{
\sum_{s,t}\sum_{s^{'},t^{'}}
<\Delta(st;s^{'}t^{'})>_{q^*}<\varepsilon(s,t)>_{q^*}\times  \nonumber
\\  &\times& <\varepsilon(s^{'},t^{'})>_{q^*} 
+ \sum_{s,t}<N{(s,t)}>_{q^{*} A Z}\times   \nonumber \\ &\times&
[<\varepsilon^2(s,t)>_{q^*}-<\varepsilon(s,t)>_{q^*}^2]\} ~ \ ,
\end{eqnarray}
where : 
\begin{eqnarray}
<\Delta(st;s^{'}t^{'})>_{q}&\equiv&<N{(s,t)}N{(s^{'},t^{'})}>_{q A Z}-\nonumber
\\&-&
<N{(s,t)}>_{q A Z}<N{(s^{'},t^{'})}>_{q A Z} \nonumber
\end{eqnarray}
and
\begin{eqnarray}
\label{eq14}
&&<N{(s,t)}N{(s^{'},t^{'})}>_{q A Z}=\nonumber \\
&&{\cal Z}_q{(s,t)}{\cal Z}_q{(s^{'},t^{'})}
\frac{{\cal Q}_q{(A-s-s^{'},Z-t-t^{'})}}{{\cal Q}_q{(A,Z)}}+\nonumber \\
&+&\delta_{ss^{'}}\delta_{tt^{'}}{\cal Z}_q{(s,t)}\frac{{\cal
Q}_q{(A-s,Z-t)}}{{\cal Q}_q{(A,Z)}} ~ \ .  
\end{eqnarray}

Fig. {\ref{fig1} shows the caloric curve for different 
values of the entropic index $q \geq 1$ in the above described 
canonical multifragmentation model.
The calculations are done for the system with $Z=79$ protons and $N=118$
neutrons ($A=197$). The free volume is : $V_f=3A/{\rho}_0\equiv A/\rho_f$, 
where $\rho_0=0.168$fm$^{-3}$. The excitation energy is : $E^*={\cal
E}_q(T,\rho_f)-{\cal E}_q(T=0,\rho_0)$. Curves $T(E^*/A)$ 
for different $q$ are very similar outside of the 'critical zone' 
of excitation energies : $E^*/A \in (2.5 - 10)$MeV. On the other hand, a  
strong sensitivity
to even tiny changes of $q$ can be seen inside of the 'critical zone'. 

This fragility of equilibrium ($q=1$) 'critical behavior' to the small 
changes  of the entropic index can be seen even
better in Fig. {\ref{fig2} which shows the specific heat vs the temperature. 
Increasing of the entropic index $q$ is associated
with both a significant sharpening of the peak in $C_V$ and an 
increase of the 'critical' temperature 
$T_C$. Since this increase of
$T_C$ is accompanied by only a small change of the total
excitation energy (see Fig. \ref{fig1}), therefore the kinetic part
in the total energy increases with $q$.
 In other words, the multifragmentation in
statistical systems with $q>1$ takes place in the hotter
environment than in the limiting equilibrium case $q=1$.
\begin{figure}[h]
\epsfig{figure=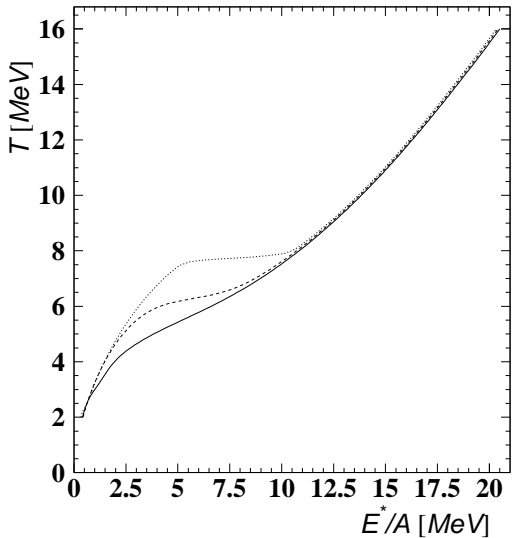,height=8cm}
\caption{The 'caloric curve' for the system with $A=197$ nucleons is plotted 
for various values of the entropic index $q$ : 1 (the solid line), 1.0005 (the 
dashed line)and 1.001 (the dotted line).}
\label{fig1}
\end{figure}
\begin{figure}[h]
\epsfig{figure=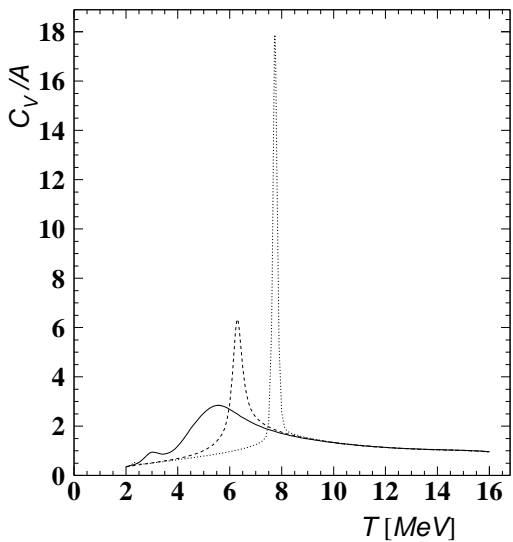,height=8cm}
\caption{The specific heat for the system with $A=197$ nucleons is plotted 
for various values of $q$ : 1 (the solid line), 1.0005 (the 
dashed line)and 1.001 (the dotted line).} 
\label{fig2}
\end{figure}

It is an open question whether the  correlations for $q\neq 1$
change the nature of the equilibrium 'phase transition'. 
Whereas the liquid-gas phase transition
 is characterized by properties of the largest cluster
\cite{stauffer}, the shattering phase transition in 
off-equilibrium systems is characterized by the 
multiplicity of fragments \cite{botet}. Fig. \ref{fig3} shows
the average multiplicity dependence of the normalized second 
factorial cumulant moment of the multiplicity distribution : 
$\gamma_2=(<m(m-1)>-<m>^2)/<m>^2$.
$\gamma_2$ is a measure of the fragment - fragment correlations and 
equals 0 for the Poisson distribution. 
One can see strong build up of multiplicity fluctuations with
increasing $q$ in the 'critical region'. This enhancement of $\gamma_2$ is
associated with the strong pic in $C_V$ as seen in Fig. \ref{fig2}.
For $q=1.001$, the maximum of $\gamma_2$
is comparable with those found in the 2D and 3D percolation systems
of comparable size \cite{bpl}. 
\begin{figure}[h]
\epsfig{figure=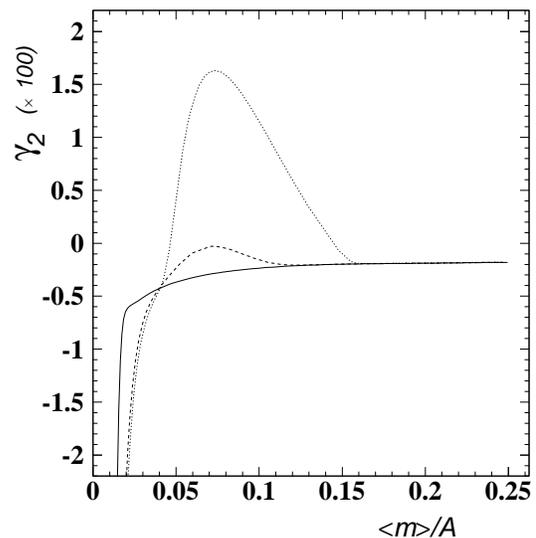,height=8cm}
\caption{The normalized second cumulant factorial moment 
of the fragment multiplicity distribution
is plotted as a function of the average multiplicity 
for various entropic indices $q$ : 1 (the solid line), 1.0005 (the 
dashed line)and 1.001 (the dotted line).} 
\label{fig3}
\end{figure}

The fragment-size distributions $dN/dA$ at $T\sim T_C$ for different $q$ values
are shown in Fig. \ref{fig4}. One can see the significant evolution of $dN/dA$
with increasing entropic index which, together with the evolution of
multiplicity distributions (see Fig. \ref{fig3}), 
illustrate the change of mechanism of the
multifragmentation. At $T=T_C$, one finds approximately power-like
fragment-size distribution for $q=1$, and the
persistence of the heaviest residue for $q>1$. The 'critical zone' for
$q>1$ is associated with the exponential fragment-size distribution. 
This resembles the critical binary fragmentation with the Gaussian dissipation
\cite{botet1} which is an off-equilibrium process.
For $T/T_C>1$ and $q>1$, the heavy residue explodes
into the large number of light fragments and the fragment-size distribution
remains exponential.

In conclusion, we have developed the generalization of canonical
multifragmentation model in TGSM. This new model 
provides an alternative way of taking into account
expected deviations from the thermodynamical equilibrium due to nonextensive
correlations in the multifragmentation process. We see the
main advantage of the proposed approach in the correct description of 
of produced fragments and in the preservation of mathematical
structure of thermodynamics. Variability of different signals of  
equilibrium phase transiton with respect to even small deviations from $q=1$, 
demonstrates that the characterization of nuclear
phase transition in terms of finite-size scaling analysis may be hazardous. 
On the other hand, signals of 'criticality' in $q$-generalized canonical 
fragmentation model are stronger, 
what in turn demonstrates the robustness of this  
'critical phenomenon'. It is worth mentioning that all considered 
characteristics change qualitatively with $q$ in the 'critical zone' of
excitation energies and one cannot exclude that order of the phase transition 
changes as well.
\begin{figure}[h]
\epsfig{figure=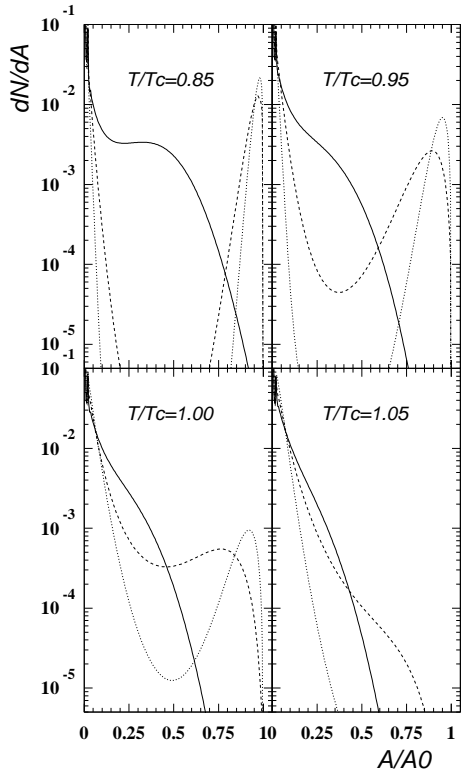,height=11cm}
\caption{The fragment-size (mass) distribution 
for the system with $A=197$ nucleons ($Z=79$ and $N=118$) is plotted 
for entropic indices and different temperatures normalized by $T_C$.}
\label{fig4}
\end{figure}
The new flexible family of fragmentation models 
obeying $q$-statistics provides a powerful tool in analyzing experimental 
data and in characterizing possible deviations from idealized equilibrium 
phase-transition picture in nuclear multifragmentation in terms of the entropic index.
This analysis is now in progress.

\vspace{1cm}
The work was supported by  the CNRS-JINR agreement No 96-28.


\end{document}